\documentclass[11pt]{article}
\usepackage[margin=1in]{geometry}
\usepackage{amsmath,amssymb}
\usepackage[dvipsnames]{xcolor}
\usepackage{tikz}
\usetikzlibrary{arrows.meta}
\definecolor{sigblue}{HTML}{0072B2}
\definecolor{ctlorange}{HTML}{E69F00}
\usepackage{graphicx}
\usepackage{array}
\usepackage{booktabs}
\usepackage{threeparttable}
\usepackage{caption}
\usepackage{placeins}

\usepackage[colorlinks=true,linkcolor=MidnightBlue,citecolor=MidnightBlue,
            urlcolor=MidnightBlue]{hyperref}
\usepackage{url}

\graphicspath{{figures/}}

\title{Collision-based logic in Lenia\\ and its composition boundary}
\author{Chakshu Gupta\\[2pt]
  {\small College of Computing, Georgia Institute of Technology}\\
  {\small \texttt{cgupta65@gatech.edu}}}
\date{}

\begin{document}
\maketitle

\begin{abstract}
Continuous cellular automata such as Lenia spontaneously produce lifelike,
self-propelling patterns, including the Orbium glider, which travels in a
straight line while pulsing through a fixed breathing cycle. Collision-based
logic,
where moving patterns compute by colliding, is established in discrete cellular
automata and continuous physical media. Within continuous cellular automata,
computation has so far been trained into the rule rather than emerging from
collisions, and whether a fixed-rule automaton like Lenia can support general
collision-based computation remains open. This paper constructs an INHIBIT gate
from collisions of the Orbium glider. Of the patterns searched across four
continuous-CA rule types, the Orbium glider is the only one shown to survive a
collision with both copies intact. A control glider deflects a signal glider off
its output line, so the output carries a signal only when no control is present.
The gate blocks across all twenty-four phases of the breathing cycle and nine
integer offsets of the control.
Two such gates in series, with one signal line and two controls, compose into an
AND-NOT chain, correct on all eight input combinations. By contrast, routing a signal beyond that single chain is
undemonstrated. A deflected signal is not restored to a fixed landing position,
and no reusable absorber for the surviving gliders was found. The immediate open question for collision-based computation in Lenia therefore
narrows from whether a gate exists to whether a deflected signal can be delivered to
a downstream gate, the next requirement for composing the gate beyond a single
straight chain.
\end{abstract}

\section{Introduction}
\label{sec:intro}

Lenia is a continuous cellular automaton. Its cells hold real values in $[0,1]$,
and a smooth growth rule updates them in small time steps, producing hundreds of
self-organising patterns, many of which move, rotate, or repair themselves like
microscopic organisms~\cite{Chan2018Lenia}. One of these patterns, the Orbium
glider, travels in a straight line while pulsing through a fixed breathing cycle.
Mobile localised patterns of this kind are the raw
material of collision-based computing, the paradigm in which information rides on
travelling localisations and is processed where they meet~\cite{Martinez2011Supercolliders}.

The paradigm has been established in two settings, neither of them a continuous
cellular automaton. In discrete cellular automata, colliding gliders implement
Boolean gates and Turing-universal machinery. Elementary rules such as Rule~54
and Rule~110 support glider reactions that encode
logic~\cite{Martinez2013SolitonECA}, and reversible Fredkin and controlled-NOT
gates have been built from glider collisions~\cite{Martinez2018LogicalGates}.
Evolutionary search has discovered gate-bearing rules
automatically~\cite{SapinBull2008Evolutionary}, and the billiard-ball cellular
automaton and soft-sphere lattice gases, whose colliding particles carry the
bits, are computation-universal~\cite{Margolus2002SoftSpheres}. In continuous physical media, the
same idea is realised with travelling waves rather than cells. Wave
fragments in the Belousov--Zhabotinsky reaction collide to form a one-bit
adder~\cite{Adamatzky2015BZAdder}, and colliding optical solitons compute in a
homogeneous nonlinear medium~\cite{Jakubowski1998StateTransform,
Steiglitz2001ManakovUniversal}. Computation has also been demonstrated inside
neural cellular automata, but there the logic is trained into the
update rule by gradient descent~\cite{Bena2025Universal, Miotti2025DiffLogic},
not produced by collisions between emergent patterns. Variants of Lenia have
been studied for adaptive agency in Particle Lenia~\cite{Horibe2023Exploring}
and open-ended evolution in Flow Lenia~\cite{Plantec2022FlowLenia}. Lenia's
capacity for
computation was raised as a question at its introduction~\cite{Chan2018Lenia},
yet no emergent glider-collision logic gate appears to have been reported in a
continuous cellular automaton, leaving open whether Lenia's gliders can
compute by colliding.

This paper constructs a logic gate and then tests how far it composes. The
signal is the Orbium glider. Of the patterns searched across four continuous-CA
rule types, it is the only one shown to keep two copies intact through their collision
(Section~\ref{sec:generality}). A control glider deflects the signal off its
output track, so a signal reaches the output only when no control is present, the
INHIBIT function. The collision is non-annihilating, and the gate blocks at
every breathing phase across nine integer
offsets of the control. Two such gates compose into an AND-NOT chain. Because the
chain passes a signal only when no control is present, a live output and deflected
survivors never coexist in it. It therefore sidesteps, rather than solves, the
disposal of surviving deflected gliders that a general circuit would have to
clear away. Beyond
that single straight chain, two capabilities remain undemonstrated. The first
is delivery, which requires turning the deflected signal onto a new track and
landing it in a downstream gate. The deflection itself already provides such a
turn, but the deflected signal is not restored to a fixed landing position,
and no restoration gadget was demonstrated. The second
is disposal, which requires absorbing the surviving deflected gliders, and no
reusable absorber was demonstrated either. The paper
claims neither a working computer nor a proof that Lenia cannot host one; it
narrows the immediate open question from whether a gate exists to whether a
deflected signal can be delivered to a downstream gate.

\section{Methods}
\label{sec:methods}

Building and testing the logic requires the Lenia substrate with its Orbium
glider, an encoding of gliders as bits and collisions as gates, and a
simulation that measures the outcome.

\begin{figure}[htbp]
  \centering
  \includegraphics[width=0.89\linewidth]{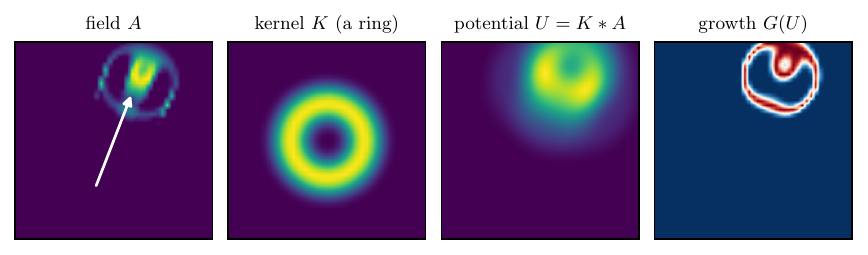}
  \caption{One update step of the Lenia rule. The field $A$
  (here the Orbium glider; brightness is the cell value, dark $0$ to bright $1$;
  the arrow marks the travel direction) is convolved with the ring kernel $K$
  (zero at the centre, peaking at half the kernel radius) to give the potential
  $U = K*A$, a local ring-average. The growth $G(U)$ (red $=+1$, grow; blue
  $=-1$, decay) is then scaled by the time step $dt$ and added to the field. The sum is clipped to $[0,1]$. The growth
  is positive at the leading edge and negative at the trailing edge, together
  advancing the glider.}
  \label{fig:substrate}
\end{figure}

\subsection{Lenia and the Orbium glider}
\label{sec:model}
Lenia evolves a real-valued field $A$ with each cell in $[0,1]$, on a square grid
in discrete time~\cite{Chan2018Lenia}. Each step first convolves the field with a
fixed radial kernel $K$ to form a potential $U = K * A$, a local weighted average
because $K$ is non-negative, normalised to unit sum, and supported within a
radius $R$ (Figure~\ref{fig:substrate}). A growth function $G$ of the potential then drives
the update. The standard rule, used here,\footnote{An asymptotic formulation
relaxes the field toward a target instead of clipping and supports its own
gliders~\cite{Davis2024NonPlatonic, Kojima2025Glider}; it is not used here.}
adds the growth $G(U)\in[-1,1]$ with time step $dt$ and clips the field,
\[
  A \leftarrow \mathrm{clip}_{[0,1]}\bigl(A + dt\,G(U)\bigr).
\]

The signal carrier is Orbium, a Lenia glider~\cite{Chan2018Lenia, Chan2020Expanded}
with a $20\times20$ seed. The rule uses kernel radius $R=13$
and time step $dt=0.1$. The kernel is a single polynomial bump with radial
profile $K(r) \propto \bigl(4r(1-r)\bigr)^4$ for relative radius
$r = \rho/R \in [0,1]$, where $\rho$ is the distance from the kernel centre;
the profile peaks at $r = \tfrac12$ and is zero for $r > 1$, so on the grid it
is a soft ring at half the kernel radius. The growth
has centre $\mu=0.15$ and width $\sigma=0.014$,
\[
  G(U) = 2\,\max\!\left(0,\, 1 - \frac{(U-\mu)^2}{9\sigma^2}\right)^{4} - 1.
\]
Under this rule, Orbium is a coherent travelling glider. Over $200$ steps it
moves in a straight line at speed $\lvert v\rvert = 0.603$ pixels per step, holds
that line with a residual scatter of $0.10$ px about a straight-line fit, varies
in mass by at most $2.4\%$, and stays localised. It also breathes, cycling
through an internal deformation with a period of $24$ steps. This breathing phase
is a nuisance variable a robust gate
must tolerate, so the experiments below sweep it across the full cycle.

\subsection{Collisions as logic}
\label{sec:collisions}
\begin{figure}[htbp]
  \centering
  \includegraphics[width=0.56\linewidth]{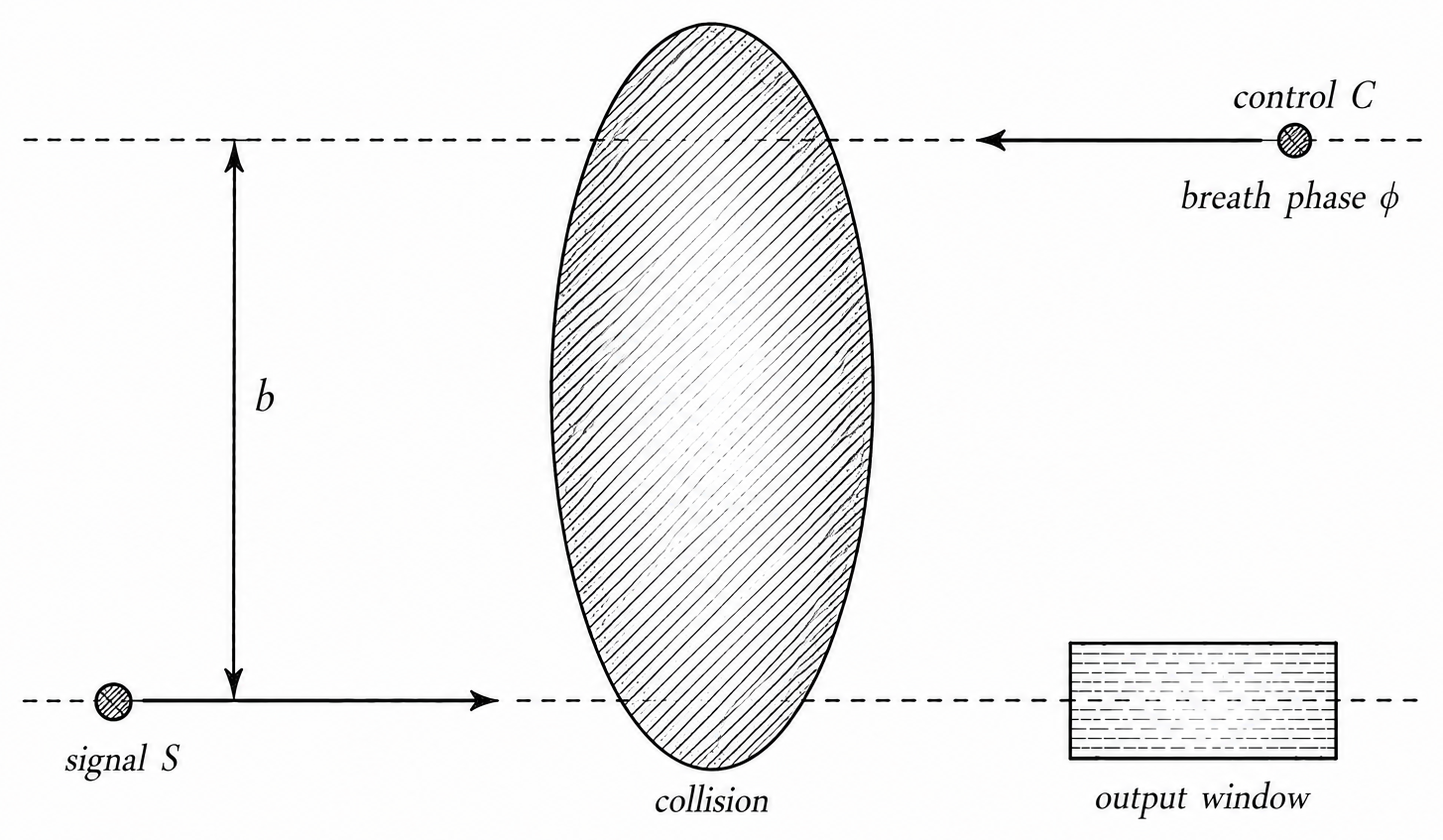}
  \caption{The collision setup, here with a head-on control. The signal $S$ and
  control $C$ are gliders on
  straight tracks through the uniform field; the control is a $D_4$ image of the
  signal, offset by the impact parameter $b$ and carrying the relative breathing
  phase $\phi$. The output bit is read at the window on the signal's outgoing track.}
  \label{fig:setup}
\end{figure}

A glider travels in a straight line across the otherwise-uniform field; there is
no fixed wiring, so a \emph{track} is just a glider's trajectory. The signal
$S$ is a glider on a track; its presence carries $1$ and its absence $0$. A
logic operation is the outcome of a collision between two gliders, so a
gate needs a second glider, the control $C$, aimed at the signal
(Figure~\ref{fig:setup}); $C$ likewise carries a bit through its presence or
absence. The control is a $D_4$ image of the signal, where the dihedral group
$D_4$ is the square's four rotations and four reflections. This group permutes
the grid cells while preserving their distances, and, since the kernel is
radial and growth and clip act cell-by-cell, the update commutes
with $D_4$. Any $D_4$ image of a glider is
therefore again a glider, its heading rotated or reflected along with its shape.
A $D_4$ image that preserves the signal's heading runs parallel to it and never
collides, so the control is drawn from the images with a different heading.
With such an image fixed, a collision is determined by the pair $(b,\phi)$,
where the impact parameter $b$ shifts the control's track transverse to its
own heading and $\phi$ is the control's relative breathing phase. For $b=0$
the two tracks meet at the field's centre. The output is read at a fixed window on
the signal's outgoing track; the window reads $1$ if a glider is present there
and $0$ if not. A signal deflected off the track is absent from the window, so
the reading is $0$ even though the glider survives elsewhere.

Whether some choice of control and offset $b$ yields a
phase-stable gate, one whose output is the same at every breathing phase $\phi$,
is settled by simulation, in two passes. A coarse survey of the collision outcome
over $(b,\phi)$ produces an operating map on which a monochromatic column marks
an offset $b$ at which the outcome does not vary across the sampled phases; a
precise sweep then evaluates a candidate gate exhaustively.
Section~\ref{sec:measurement} details the placement and rotation of
collisions used in both passes.

\subsection{Simulation and measurement}
\label{sec:measurement}
The field $A$ is evolved on a square periodic grid in double precision, because
colliding gliders are sensitive to rounding. Each step evaluates the kernel
convolution $K * A$ as $\mathcal{F}^{-1}(\hat{K}\,\mathcal{F}A)$, where
$\mathcal{F}$ is the Fourier transform and $\hat{K} = \mathcal{F}K$ is
precomputed once.

A collision is classified from its \emph{blobs}, the connected components of
$\{A > 0.10\}$ whose mass is at least half a lone glider's, and from the
run's total field mass. The run is flagged as an interaction if the
total field mass ever departs by more than $15\%$ from twice a lone glider's
mass, or if a blob sample taken every four steps ever finds fewer than two
blobs. The second condition catches mass-conserving events such as a
reflection, in which the two gliders form a single blob near closest approach.
At the run's end, the label follows from the blob count and the mass ratio
$r$, defined as the total field mass divided by twice a lone glider's mass. A run is
labelled \emph{annihilate} when $r<0.10$, \emph{explode} when $r>1.80$,
\emph{merge} when a single blob remains and $r\in(0.30,\,1.70)$, and
\emph{survive} or \emph{miss} when two or more blobs remain and
$r\in(0.70,\,1.70)$; a fired interaction flag makes the run a survive, an
unfired flag a miss. Any run matching none of these
conditions is labelled \emph{residual}. After a fixed-length run, the bit
$\text{out}$ is computed at the output window from the last frame and the
frame $20$ steps earlier; the window is a disc of radius $20$ cells centred at
the signal's expected arrival point on its outgoing track. The bit is $1$ when
a blob of mass at least $0.60$ of a lone glider's lies within the window with
velocity, measured over the previous $20$ steps, within $35^\circ$ of the
signal's direction. Otherwise the bit is $0$.

A coarse survey at breathing phases $\phi = 0, 3, \dots, 21$ and integer
offsets $b = 0, \dots, 24$ produces the operating map, on which each sampled
$(b,\phi)$ is coloured by its label. A column of one colour marks an offset
$b$ at which the label does not vary across the sampled phases. The precise
sweep then evaluates a candidate gate by running the control-present collision
at every breathing phase $\phi = 0,1,\dots,23$ and every integer offset in the
gate's chosen range, on a $361\times361$ grid sized to keep the gliders and
any collision products clear of the periodic boundary. For the gate reported
below, this range is $b = 0,1,\dots,13$. The control for phase $\phi$ is the glider's seed evolved
$\phi$ steps to set its breathing phase, then reoriented into the collision
geometry by an exact $D_4$ permutation of its cells rather than an
interpolation. Because a glider's target centre of mass may fall between
integer cells, the signal and the control are each placed by a Fourier-domain
shift, a linear phase ramp on their transforms, removing the up-to-half-cell
error that plain integer placement would leave in each coordinate. The robust
window is $\{\,b : \text{out}=0 \text{ for every phase }\phi\,\}$, the offsets
at which the control blocks the signal at every phase. This procedure draws no
random numbers, and the counts and windows below reproduce across
reruns.\footnote{The simulation and analysis code is available at
\url{https://github.com/ChakshuGupta13/lab}.}

\section{Results}
\label{sec:results}

\subsection{The collision as a switch}
\label{sec:phenomenology}
\begin{figure}[htbp]
  \centering
  \includegraphics[width=0.86\linewidth]{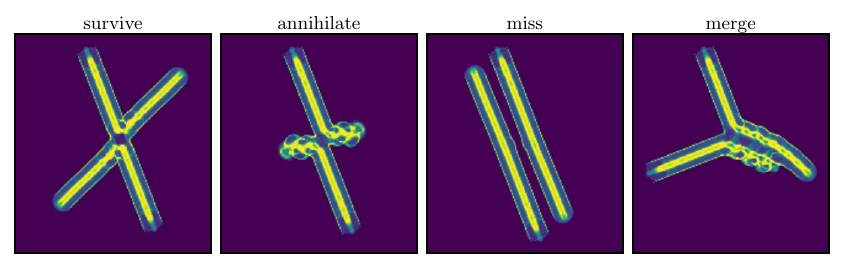}
  \caption{The four collision outcomes as max-over-time projections of the
    field. Head-on Orbium collisions give survival at $b=0$, annihilation at
    $b=7$, and a miss at $b=22$; the fourth outcome, a merger, appears at $b=0$
    in the perpendicular geometry used in the gate (Section~\ref{sec:gate}), and
    also head-on (Figure~\ref{fig:opmap}). Each panel is one
    simulator run at $\phi=0$.}
  \label{fig:outcomes}
\end{figure}

A logic gate built from collisions needs gliders that survive them. Of the
eighteen soliton families catalogued for Lenia, only Orbium's family, the
Orbidae, survives collisions; the others lose self-organisation~\cite{Chan2020Expanded}. The gate also
needs a fixed collision outcome. When two Orbia meet head-on, the result is not a
single fixed reaction. As the impact parameter $b$ is varied, the outcome runs
through distinct classes (Figure~\ref{fig:outcomes}). At some
offsets, both gliders survive; at others, the pair annihilates, leaving an empty
field.

\begin{figure}[htbp]
  \centering
  \includegraphics[width=0.67\linewidth]{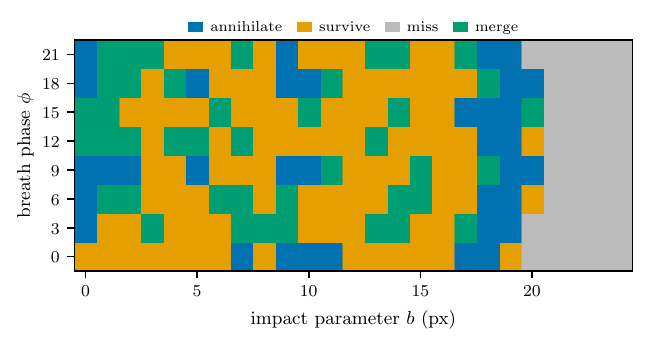}
  \caption{Head-on Orbium collision outcome, plotted horizontally over the
    impact parameter $b$ and vertically over eight relative breathing phases
    $\phi$ spanning the cycle. The miss region beyond $b=20$ is uniform across
    phases, and survival is the most common outcome. Within $b\le20$, only two
    of the twenty-one columns hold one outcome across all eight phases; in the
    rest, survival, annihilation and merger interleave. Here $b$ alone therefore
    does not fix the head-on outcome.}
  \label{fig:opmap}
\end{figure}

The head-on outcome is not a function of $b$ alone.
Figure~\ref{fig:opmap} maps it over the impact parameter $b$, sampled at integer
offsets from $0$ to $24$, and the relative breathing phase $\phi$, at eight
values spanning the cycle. Survival is the most common outcome, filling $88$ of
the $200$ cells. Beyond $b=20$ the gliders miss at all eight phases, a uniform
band. At smaller offsets, survival, annihilation and merger interleave. Of
the twenty-one columns with $b\le20$ only two are phase-uniform, both giving
survival, and the two are not adjacent, so no band of neighbouring offsets in
this range holds one outcome at every phase. The next section builds and
certifies a geometry where a perpendicular $D_4$ partner blocks a glider from
the output window across a contiguous run of offsets at every phase.

\subsection{An INHIBIT gate by deflection}
\label{sec:gate}
\begin{figure}[htbp]
  \centering
  \includegraphics[width=0.44\linewidth]{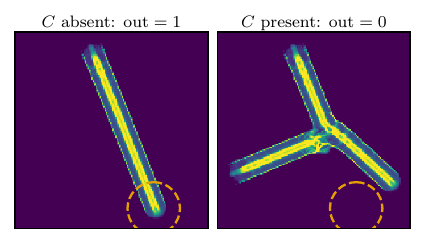}
  \caption{The INHIBIT gate, shown as the max-over-time projection of the
    field so that each bright streak is a glider's whole path. The signal enters
    from the top toward the dashed output window. With the control absent, the
    signal reaches the window, giving $\text{out}=1$ (left); with the control
    present, the collision blocks the signal, giving
    $\text{out}=0$ (right).}
  \label{fig:gate}
\end{figure}

The gate computes $\text{out} = S \wedge \neg C$ from two gliders. A signal
glider $S$ travels toward a downstream output window; a control glider $C$, the
$90^{\circ}$-rotated $D_4$ partner of the signal, crosses $S$'s track
perpendicularly at impact parameter $b$, unlike the preceding head-on switch.
When $C$ is present the collision blocks the signal, and
the output, evaluated when the signal is due at the window, reads $0$; when
$C$ is absent, $S$ travels straight through, and the output reads
$1$ (Figure~\ref{fig:gate}). The block is non-annihilating. At impact parameter
$b = 5$ (Figure~\ref{fig:gate}), a live glider survives in most breathing phases;
in the rest, two gliders' worth of mass survives, either as a single lump or a
separated pair. Because the block
deflects the surviving mass off the signal's track, that mass is clear of the
output window at readout; the perpendicular control never reaches it either, and
with neither glider present nothing does. Across the four input combinations,
only $S=1, C=0$ outputs $1$, giving the INHIBIT table.

\begin{figure}[htbp]
  \centering
  \includegraphics[width=0.5\linewidth]{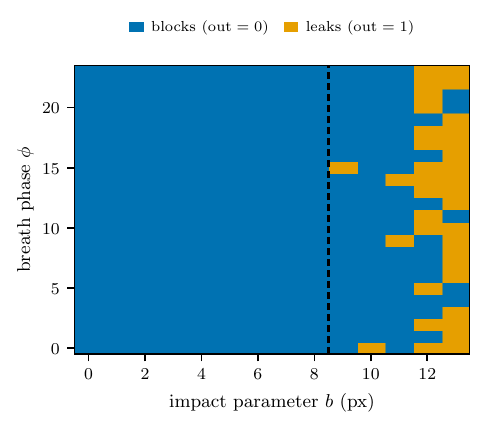}
  \caption{Gate robustness, showing whether the gate blocks the signal over
    the impact parameter $b$ (horizontal) and all twenty-four breathing phases
    (vertical). The gate blocks at every phase across the integer offsets
    $b=0$ to $8$, left of the dashed line; the first integer leak is at $b=9$.}
  \label{fig:robust}
\end{figure}

Since the control's breathing phase is not synchronised, the gate must block at
whatever phase the control arrives. A sweep over all
twenty-four breathing phases against the impact parameter $b$
(Figure~\ref{fig:robust}) shows the gate blocking at every phase across the nine
integer offsets $b = 0$ to $8$; the first integer offset at which it leaks,
reading $1$ instead of $0$, is $b = 9$, and over the range $b = 0$ to
$13$ the gate blocks at $89\%$ of the $336$ points. The integer
samples miss a few isolated leaks between them, however. Under a finer sweep of
real-valued offsets, the gate first leaks near $b = 4$, at scattered non-integer
offsets, so the integer window overstates the tolerance
to continuous placement.

\subsection{Two gates compose in series}
\label{sec:cascade}
Two gates placed in series compute $\text{out} = S \wedge \neg C_1 \wedge \neg
C_2$. The signal passes the first gate's output window only if $C_1$ is absent,
then the second only if $C_2$ is absent. Across all eight combinations of the
three inputs, the chain produces the correct output, and
it does so at four inter-gate spacings of about $37$, $49$, $67$, and $85$
pixels, so the stages need not be a fixed distance apart. This is input-to-output composition of two
collision gates.

The chain's composition is restricted. This is a real limitation rather
than a detail. In a general circuit, some gate would output a $1$ while a live
survivor from an upstream collision is still on the grid. The AND-NOT chain never
does, because it emits a $1$ only when neither control is present. The chain
therefore never requires survivor disposal, but general composition does. The next
section addresses that disposal.

\subsection{The composition boundary}
\label{sec:boundary}
Beyond a single straight chain, a general circuit needs more. It must turn a
signal, fan one signal into two, let two signals cross, deliver a turned
signal into the next gate, and absorb the live survivors that deflection
leaves behind. Table~\ref{tab:boundary}
summarises the results. The signal carrier, the gate, and a turn work in
isolation, as does a crossing of two signals on perpendicular tracks, provided the
signals are separated by about sixty steps in time to avoid a collision. A one-into-two fan-out, in
which a counter-propagating helper scatters the signal into two survivors, works
only partly, since their onward tracks are unverified. The last two, turn-to-gate
delivery and survivor absorption, were not demonstrated. Delivery was not tested
end to end, and no clean absorber was found, so the
construction goes no further than a single chain.

The first undemonstrated capability is delivery of a turned signal to a
downstream gate. Orbium survives the collision as a live glider, but the turn
leaves its position and heading uncorrected. The exit
heading's circular standard deviation ranges from about eight to fourteen degrees
across the measured control offsets. On one route
followed downstream, the spread of the landing point widens with distance,
reaching about sixteen pixels from the mean at about seventy pixels along it,
with no correction observed over that route. Whether a downstream gate can still
act on a signal displaced this far was not measured. The gate's tolerance
to a displaced, reoriented incoming signal was never swept, and no turn-to-gate
route was run end to end. Arranging for the signal and its turning partner to
arrive in a fixed relative phase narrows an isolated turn's landing spread, but whether
that narrowing persists under the coupled timing of a multi-stage chain was not tested.
One way to close the delivery gap is a gadget that restores a signal's position and
heading, a reflector or waveguide; none was constructed here.

The second undemonstrated capability is a clean absorber for the live survivors that
deflection produces. Two families were searched (Table~\ref{tab:absorb}). A
mutual-annihilation eater, a partner that drives the total mass to near zero, was
sought at eight breathing phases over the impact-parameter sweep, for each of six
collision geometries, namely the head-on, the two perpendiculars, the two axis
reflections, and the main diagonal. Clean annihilation occurs but is phase-fragile,
reaching at best six of the eight phases at any single offset and never all
eight, so no phase-robust eater was found in this family; and since the eater is
consumed in the annihilation, even a phase-robust one would absorb a survivor
only once. A static still-life that could act as a persistent absorber was
sought by relaxing twelve simple seeds, all of which die, and by a
fixed-point search from eighteen starts, none of which yields a still-life that survives on a large grid.
Neither family therefore provides a reusable absorber. Untested cases include a catalytic eater that survives the absorption, a
stationary breathing pattern, and multi-body or phase-scheduled absorbers. The finding is that no reusable absorber was found in the families
searched, not that none exists.

Together the two gaps leave universal computation open in a specific, measured
sense. No depth-two construction that routes a turned signal or disposes
of a survivor has been demonstrated, and no per-primitive impossibility has been
proved. Universality is therefore undetermined.

\begin{table}[htbp]
  \centering
  \footnotesize
  \begin{tabular}{|l|l|p{0.56\linewidth}|}
    \hline
    \textbf{Primitive} & \textbf{Status} & \textbf{Constraint} \\
    \hline
    Signal carrier     & Works     & A single glider carries a bit along a track. \\
    \hline
    INHIBIT gate       & Works     & The control must fall within a leak-free band about four pixels wide. \\
    \hline
    Fan-out            & Partial   & A helper yields two survivors, but their tracks are unverified. \\
    \hline
    Turn               & Works     & The exit heading has a circular standard deviation of $8$--$14^{\circ}$ across offsets; the landing spreads by about $16$ px at $70$ px downstream. \\
    \hline
    Crossing           & Works     & The two signals must be separated by about $60$ steps in time. \\
    \hline
    Turn-to-gate delivery & Untested  & The turn leaves the landing point uncorrected, and the downstream tolerance was not swept. \\
    \hline
    Survivor absorber  & Not found & No phase-robust eater or still-life was found. \\
    \hline
  \end{tabular}
  \caption{The depth-one primitive library. Four primitives function in
    isolation and a fan-out only partly; two capabilities remain undemonstrated,
    so the library supports only the reported straight chain.}
  \label{tab:boundary}
\end{table}

\begin{table}[htbp]
  \centering
  \footnotesize
  \begin{tabular}{|>{\raggedright\arraybackslash}p{0.17\linewidth}|>{\raggedright\arraybackslash}p{0.33\linewidth}|>{\raggedright\arraybackslash}p{0.40\linewidth}|}
    \hline
    \textbf{Family} & \textbf{Search space} & \textbf{Result} \\
    \hline
    Collision eater & 6 geometries $\times$ 8 phases $\times$ 13 offsets & Annihilation is clean in at best 6 of the 8 phases, never all. \\
    \hline
    Static still-life & 12 seeds + 18 fixed-point starts & No still-life emerged, and every seed died. \\
    \hline
  \end{tabular}
  \caption{Absorber search. Two families were searched against a clean-removal
    criterion (final mass ratio ${<}0.10$ for the eater; a non-moving,
    mass-stable pattern for the still-life); neither yields a reusable absorber.}
  \label{tab:absorb}
\end{table}

\FloatBarrier

\subsection{Generality of the deflection prerequisite}
\label{sec:generality}

The deflection prerequisite is that a two-glider collision leave both
gliders, each within $\pm30\%$ of a single glider's mass.
Table~\ref{tab:generality} tests it across four continuous-CA rule types. The
gate needs gliders
robust enough to survive a collision, and clean two-glider survival is a test
of that robustness. Of the substrates searched, only
Orbium-in-standard-Lenia passes it. In the others the gliders lose coherence, the
collision is not reproducible, or no isolated glider forms at
all.

Within standard Lenia, the Orbium control and ten further species from the
expanded Lenia catalogue~\cite{Chan2020Expanded} are tested. Each is collided at
impact parameters $b=0,2,\ldots,24$ under six partner geometries,
namely the head-on, the two perpendiculars, the two axis reflections, and the
main diagonal. This sweep is coarser and wider than the gate's certifying one,
without its scan over breathing phase. One of the ten, Pyroscutium, is
tested with a Gaussian growth function, which replaces its native polynomial
growth and yields a stable glider.
Another, Quadrium vagus, is nearly stationary, so it is unsuited to a collision
test and is set aside, leaving nine species. None of
the nine yields a clean collision, while the
Orbium control does. An asymptotic-Lenia
glider~\cite{Davis2024NonPlatonic}, from a different rule type, is collided with a
copy of itself under the same sweep, yielding none either.

For SmoothLife~\cite{Rafler2011SmoothLife}, the survey had no published glider
seed to test. Instead, a search across seven parameter settings evolved random
initial conditions for $600$ steps, then copied thirty-five emergent
patterns onto an empty grid; none proved to be a vacuum-stable isolated
glider. With no
glider to collide, SmoothLife is recorded as untested for the prerequisite rather
than excluded from the survey.

Glaberish~\cite{Davis2022Glaberish} is implemented as its canonical s613 rule,
a state-dependent genesis--persistence update on the three-ring kernel of the
Lenia species Hydrogeminium. It reproduces the published behaviour, remodelling
continuously where Hydrogeminium settles to a static pattern. It does produce
emergent gliders, but collisions between them are not reproducible on this pipeline.
Byte-identical reruns diverge to order-one field differences within a few hundred
steps, because the backend's floating-point reductions are not bitwise
deterministic and s613's dynamics amplify that noise while Orbium's keep it
negligible. A reliable
survival count therefore cannot be taken for s613, and whether it meets the
prerequisite is left undetermined; the demonstrated gate relies on Orbium's
reproducibility instead.

\begin{table}[htbp]
  \centering
  \footnotesize
  \begin{tabular}{|l|l|c|l|}
    \hline
    \textbf{Substrate} & \textbf{Configurations} & \textbf{Clean collisions} & \textbf{Finding} \\
    \hline
    \multicolumn{4}{|l|}{\textit{Standard Lenia}} \\
    \hline
    \quad Orbium (control)            & $6\times13=78$         & 24  & Met \\
    \hline
    \quad Nine other species          & $9\times6\times13=702$ & 0   & Unmet \\
    \hline
    Asymptotic Lenia                  & $6\times13=78$         & 0   & Unmet \\
    \hline
    SmoothLife (7 parameter variants) & 35 patches             & N/A & No isolated glider \\
    \hline
    Glaberish (s613)                  & Gliders found          & N/A & Non-reproducible \\
    \hline
  \end{tabular}
  \caption{Generality of the deflection prerequisite. A collision is
    \emph{clean} when it leaves both gliders, each within $\pm30\%$ of a
    single glider's mass; the table counts the clean collisions found over the
    sweep, marking N/A where no glider forms or its collisions are not
    reproducible. Only Orbium yields any. The nine other species are
    Paraptera, Hexacaudopteryx, Pyroscutium, Hydrogeminium (3-ring),
    Heptapteryx (3-ring), Kronium (2-ring), Astrium (4-ring), Rotorbium, and
    Gyrogeminium (3-ring); the SmoothLife variants are the canonical SmoothLife
    parameters and six other points of the parameter cube.}
  \label{tab:generality}
\end{table}

\section{Discussion}
\label{sec:discussion}

A gate exists in Lenia; what it cannot yet do is compose into a general circuit.
The gate is phase-robust across nine integer offsets, and series chaining works
because a chain carries its value on the untouched signal and never has to
deliver a turned one. Routing a signal to a gate off its straight path, as a
general circuit does, turns the signal. The turn leaves such a signal's position and
heading uncorrected, and whether it then reaches the narrow window of the next
gate is untested. This is not specific to the deflection gate; any layout that
must place a turned signal precisely faces the same gap. Survivor
absorption is a second gap, and no reusable absorber for it was found here.

The comparison of Lenia with the substrates where collision-based
computing already works is instructive rather than favourable. Each of them has a means of keeping a signal usable after it
interacts. Discrete cellular automata run on a lattice, so a glider's position and
phase take known discrete values. The
billiard-ball cellular automaton and the soft-sphere lattice gases hold their
signals on fixed
paths, routing them with mirrors or, in the momentum-conserving variants, with
streams of particles~\cite{Margolus2002SoftSpheres}. In the Belousov--Zhabotinsky
medium, a one-bit adder confines the colliding wave fragments to engineered
channels, so that geometry holds each signal on its
track~\cite{Adamatzky2015BZAdder}; the excitations carry Boolean values by their
presence, exactly as the gliders here do. Optical-soliton
computers instead encode a value in a soliton's complex state, which each
collision transforms; making the computation composable takes explicit machinery, namely
restoration pulses that reset the state and a second signal speed for
routing~\cite{Steiglitz2001ManakovUniversal}. Each of
these constructions keeps a signal on a known course after a collision, whether by
quantising a glider's position to a lattice, holding it on a fixed path, or
restoring and routing a soliton; free-space Lenia restores a glider's shape but not its
position or heading. The Lenia gate here has a working encoding, a value carried
as the presence of a localisation, but no demonstrated means of keeping that
value on course. It chains in series, while composition beyond that remains
untested.

The deflection collision the gate depends on is itself a rare one; no other
configuration searched was shown to meet the deflection prerequisite
(Section~\ref{sec:generality}). A prior observation that of Lenia's
eighteen soliton families only the Orbidae, Orbium's family, collide and survive,
while the others lose self-organisation on contact~\cite{Chan2020Expanded}, aligns
with the finding here that among the standard-Lenia species tested only Orbium
meets the deflection prerequisite, a stricter screen than survival alone.
Published glider configurations the search here left unresolved, including the
published SmoothLife glider~\cite{Rafler2011SmoothLife} and the published
Glaberish gliders~\cite{Davis2022Glaberish}, are not ruled out.

The result's limits are that only series composition was demonstrated
and that the enabling collision was confirmed only for Orbium among the configurations
searched.

\subsection{Future directions}
\label{sec:future}
One natural candidate for delivering a turned signal onward is a restoration
gadget, a structure that returns a turned glider to a known position and
heading so that it lands inside the next gate's window. This gadget and the
missing absorber could both be persistent structures, patterns that sit in place and
act on a passing glider. Such structures, not systematically searched for here,
are candidates. A stationary
breathing pattern, which oscillates in place without translating and is the
natural Lenia analogue of a still-life, could stand in a glider's path and
reflect or absorb one, though the period-one fixed-point search used above cannot
represent it. A catalytic eater that survives the absorption it
performs would be a reusable absorber, though the clean-removal criterion used
above did not screen for that survival. Beyond standard Lenia, a mass-conserving
variant, where localised patterns persist rather than dissipate, is a likelier
substrate for the restoration gadget, though its conservation law rules out a
mass-destroying absorber~\cite{Plantec2022FlowLenia}.
Finally, a
constant-one source realised as a glider gun, and an analysis of whether
per-stage timing tolerances compose at scale, are among the further requirements
for a full construction.

\section{Conclusion}
\label{sec:conclusion}
The lifelike gliders of a continuous cellular automaton can compute by collision.
In standard Lenia, collisions of the Orbium glider implement an INHIBIT gate that
is robust to the glider's shape oscillation, and two such gates compose in series.
The gate is emergent, arising from collisions between the automaton's own
patterns rather than from logic trained into its update rule. The enabling
collision was confirmed only for Orbium among the configurations searched, and
general-circuit composition was not achieved. The first gap is delivering a turned
signal to a downstream gate. The turn leaves the signal's position and heading
uncorrected, so whether the signal reaches the gate is untested. The second gap is
absorbing the survivors that a collision leaves behind, for which no reusable
absorber was found. Each established substrate discussed above has a
means of keeping a signal on a known course after a collision, so a persistent
structure that restores a turned glider's position and heading would let Lenia do
the same. A reusable absorber is another persistent structure, so the two
searches overlap.

\bibliographystyle{alpha}
\begingroup
\raggedright
\bibliography{references}
\endgroup

\end{document}